\documentclass[pra,showpacs,twocolumn,superscriptaddress]{revtex4}
\usepackage{amssymb}
\usepackage{graphicx}
\usepackage{amsmath}
\usepackage{amsbsy}
\usepackage{amsthm}
\usepackage{epsfig}
\usepackage{color}

\newcommand{\ket}[1]{\ensuremath{|#1\rangle}}

\begin{document}

\title{Photonic Multiqubit States from a Single Atom}
\author{Ying Li}
\affiliation{Centre for Quantum Technologies, National University of Singapore, Singapore 117543}
\author{Leandro Aolita}
\affiliation{ICFO-Institut de Ci\`encies Fot\`oniques, Parc Mediterrani
 de la Tecnologia, 08860 Castelldefels (Barcelona), Spain}
\author{L. C. Kwek}
\affiliation{Centre for Quantum Technologies, National University of Singapore, Singapore 117543}
\affiliation{Institute of Advanced Studies (IAS), Nanyang Technological University, Singapore 639673}
\affiliation{National Institute of Education, Nanyang Technological University, Singapore 637616}

\begin{abstract}
We propose a protocol for the creation of photonic  Greenberger-Horne-Zeilinger and linear cluster states emitted from a single atom---or ion---coupled to an optical cavity field.
The method is based on laser pulses with different polarizations and exploits the atomic transition amplitudes to state-selectively achieve the desired transitions.  The scheme lies within reach of current technology.\end{abstract}

\maketitle

\section{Introduction}
The experimental effort devoted worldwide to the production and coherent manipulation of genuinely multiparticle-entangled states over the last decade has been  tremendous \cite{BowmeesterPanWitnessesZhao, Rauschenbeutel, Sackett, RoosHaeffner, Leibfried, kieselwalther1walther2, Lu, ChenprevedelVallonececcarelli}.
The main motivations behind this effort are arguably the potential applications of the Greenberger-Horne-Zeilinger (GHZ) \cite{GHZ} and cluster~\cite{RausBrie} states. The former can be considered as simple models of the celebrated \textit{gedanken} Schr\"odinger-cat states  \cite{Leibfried, Lu}, are crucial for quantum communication and cryptography problems \cite{GHZuse}, and have been found useful in metrology \cite{Giovannetti} as well as in high-precision spectroscopy \cite{Bollinger}.
The latter are massively-entangled states that make one of the main paradigms of quantum computation possible, namely, the measurement-based one-way approach \cite{RausBrie}.
There, computation proceeds by a sequence of adaptive one-qubit measurements on the cluster, consuming cluster-state entanglement as the main resource.

As a physical platform for the transmission of quantum information without significant noise, photons are the natural choice.
In addition, photonic platforms have potential for quantum information processing since all-optical models for quantum computing using only linear-optical devices, single-photon sources and detectors exist \cite{klm}.
Furthermore, in linear-optical setups, both photonic GHZ \cite{BowmeesterPanWitnessesZhao, Lu} and cluster \cite{kieselwalther1walther2, Lu, ChenprevedelVallonececcarelli} states have been demonstrated in proof-of-principle experiments with up to six photons.
However, in these setups, photon-pair generation is highly inefficient, and the entangling gates necessary to fuse these pairs into larger multi-qubit states are in addition intrinsically probabilistic. This poses a fundamental obstacle to the scaling to large numbers of particles.

On the other hand, atom-cavity systems make excellent single-photon-single-atom interfaces \cite{Rauschenbeutel, McKeever, Keller, Bochmann, Wilk, Weber, Barros}. High-efficiency single photons emitted in a predetermined spatiotemporal mode, from single neutral $\mathrm{Rb}$ \cite{Bochmann} and $\mathrm{Cs}$ \cite{McKeever} atoms, and even trapped $\mathrm{Ca}^{+}$ ions \cite{Keller, Barros}, inside an optical cavity, have been realized.
Furthermore, with similar experimental setups, single-photon-single-atom and single-photon-single-photon entanglements have been successfully demonstrated \cite{Blinov,Volz,Wilk, Weber}.

In this article we propose a family of protocols for the creation of photonic GHZ and linear cluster states emitted from a single atom -- or ion -- coupled to an optical cavity field. These protocols are based on laser pulses with different polarizations and exploit the atomic natural dipole-transition elements to state-selectively achieve the desired transitions. The methods are in principle deterministic. However, in practice the overall efficiency is never unity. We provide a detailed analysis of the sources of imperfections and show that cavity photon-emission efficiencies close to 70$\%$ per photon are feasible. The procedures are illustrated with $^{87}\mathrm{Rb}$ and $^{40}\mathrm{Ca} $ atoms as examples, respectively with and without hyperfine structure, and for whom the state-of-the-art technology is in an extremely advanced stage \cite{RoosHaeffner, Keller, Barros,Volz, Bochmann,Wilk, Weber}. Their extensions to other alkali-metal or alkaline-earth-metal species are straightforward.

\par The paper is organized as follows. In Sec. \ref{Prot} we present our ideas in abstract terms. In Secs. \ref{SecCa} and \ref{SecRb}
we describe concrete experimental procedures to implement the proposed ideas with $^{40}\mathrm{Ca} $ and $^{87}\mathrm{Rb}$ atoms, respectively. We leave an analysis of the technical details common to both implementations for Sec. \ref{TecDet}, and we devote Sec. \ref{Feas} for an assessment of the experimental feasibility with current technology and some discussions. Finally, Sec. \ref{Conclusion} contains our concluding remarks.
\section{The Protocol}
\label{Prot}
 A neutral (or ionized) atom is optically (electrically) confined inside a high-finesse optical cavity \cite{McKeever, Keller, Barros, Bochmann, Wilk, Weber}, with whose field the atom is strongly coupled. The atom is excited by laser-pulse sequences that propagate perpendicularly to the cavity axis.  One of the cavity mirrors is partially transmissive and the well-defined photonic output mode through it provides the dominant channel of atomic decay.  Repeated application of these pump  sequences produces trains of photons that are collected at the cavity output by an optical fiber, through which they propagate with the desired multiqubit states in their polarization degree of freedom.

Each of the above-mentioned pulse sequences is designed to drive either of the following state-transformations on the atom-cavity system:
\begin{subequations}
\label{GHZLC}
\begin{align}
\label{TGHZ}
{T_{GHZ}}:&|\pm\rangle\rightarrow\pm|\pm\rangle|\sigma^{\pm}\rangle,\ \  \text{or}\\
\label{TLC}
{T_{LC}}:&|\pm\rangle\rightarrow\frac{1}{\sqrt{2}}(\pm\left\vert
+\right\rangle \left\vert \sigma ^{+}\right\rangle-\left\vert
-\right\rangle \left\vert \sigma^{-}\right\rangle ).
\end{align}
\end{subequations}
Here, kets $|+\rangle$ and $|-\rangle$ stand for two long-lived atomic states in which the atomic $z$ computational basis is encoded.
$|\sigma^{+}\rangle$ and $|\sigma^-\rangle$ in turn denote the right and left circularly-polarized states, respectively, of the photon emitted in each sequence, which constitute the photonic $z$ computational states.
Transformations \eqref{GHZLC} are called isometries, mapping the atomic-qubit Hilbert space into the two-qubit atomic-photonic one.
Isometries for the sequential creation of multiqubit states were studied in Ref. \cite{Schoen} in general terms. 

\par In Appendix \ref{Transformations} we show explicitly how the repeated application of transformations \eqref{TGHZ} or \eqref{TLC}, respectively, lead to $N$-photonic-qubit GHZ  \cite{GHZ} or linear cluster \cite{RausBrie} states. In both cases the protocol consists first of the successive application of transformations \eqref{TGHZ} or \eqref{TLC}, respectively,  $N$ times. This already generates the desired multiqubit states but in the hybrid atom-$N$-emitted-photons system. Then, to decouple the atom from the state, a projective measurement  is applied to it. Naturally, such measurement is most efficiently done by taking advantage of the atomic coupling with the cavity photons. So, in both cases the atom is finally measured in the corresponding appropriate basis via a further excitation and subsequent measurement of the last emitted photon (see Appendix \ref{Transformations}).

\section{Implementation with $^{40}\mathrm{Ca}$}
\label{SecCa}
\begin{figure}
\begin{center}
\includegraphics[width=1\linewidth]{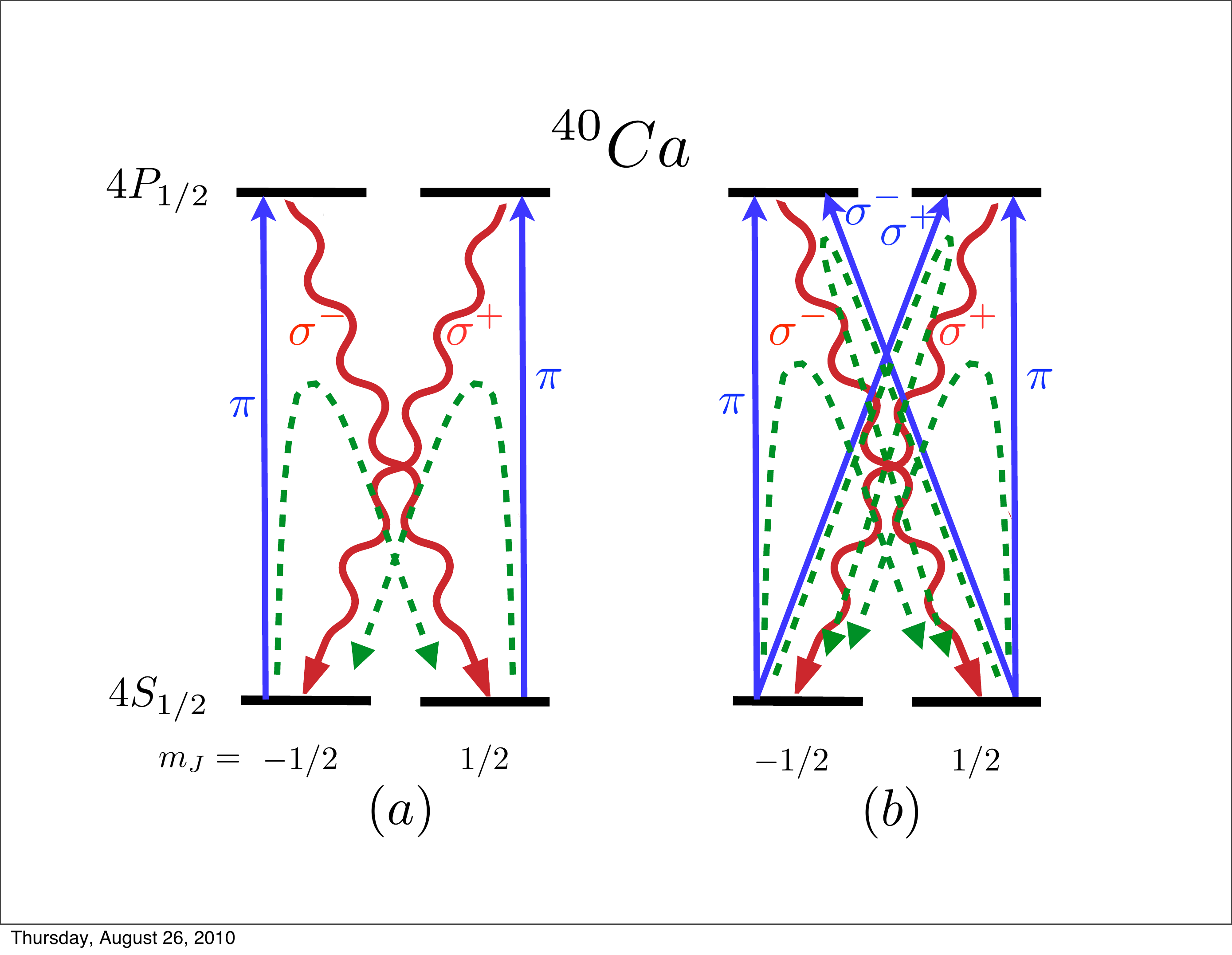}
\caption{ \label{Ca} (Color online.) Relevant fine level structure of $^{40}\mathrm{Ca}$.
Only photons (red wavy lines) with $\sigma^+$ or $\sigma^-$ circular polarizations can decay into the cavity (resonant with $4S_{1/2}\leftrightarrow4P_{1/2}$).
$(a)$ A $\pi$-polarized laser (blue lines) resonantly excites both ground-state sublevels $\ket{4S_{1/2},m_J=\pm 1/2}$ to the $4P_{1/2}$ manifold.
The total effective process including spontaneous photon emission is represented by green dashed lines, and its successive repetition creates a GHZ state.
$(b)$ Same as $(a)$ but with the laser possessing also a $\sigma^+/\sigma^-$-polarization component.
Two effective processes are driven simultaneously by the same pulse, and its repetition generates a linear cluster state.}
\end{center}
\end{figure}

This isotope does not feature hyperfine structure (see Fig. \ref{Ca}).
The Zeeman sublevels of the $S_{1/2}$ ground state encode the atomic qubit: $|\pm\rangle \equiv|4S_{1/2},m_J=\pm 1/2\rangle $, where the quantization direction is that of the cavity axis.
Both the cavity mode and a monochromatic pump laser are in resonance with the dipole transition $4S_{1/2}\leftrightarrow 4P_{1/2}$.
The laser is linearly-polarized either perpendicular to the cavity axis, decomposing into two equal components of $\sigma^+$ and $\sigma^-$ polarizations, or parallel to it.
Photons with the former polarization, $\sigma^+/\sigma^-$, can only either absorb from or deliver to the atom one quantum $m_J$ of angular momentum.
Photons with the latter polarization, $\pi$, necessarily maintain $m_J$ unchanged. Transformations \eqref{GHZLC} can both be realized with a \textit{single laser pulse}.

For the case of \eqref{TGHZ}, the pump laser is $\pi$-polarized and drives the excitations $\left\vert S_{1/2},m_J=\pm1/2\right\rangle \rightarrow \left\vert P_{1/2},m_J=\pm1/2\right\rangle$.
The cavity in turn supports only $\sigma^+/\sigma^-$ polarizations, because photons propagating along the cavity axis cannot carry $\pi$ polarization.
Thus atomic decay takes place only through the transitions $\left\vert P_{1/2},m_J=\pm1/2\right\rangle \rightarrow\pm\left\vert S_{1/2},m_J=\mp1/2\right\rangle$ (the sign factor coming from the Clebsch-Gordan coefficients), accompanied by the corresponding emission of a $\sigma^{\pm}$ photon into the cavity. Altogether, the ground-state sublevels transform as $|S_{1/2},m_J=\pm1/2\rangle\rightarrow\pm|S_{1/2},m_J=\mp1/2\rangle |\sigma^{\pm}\rangle $.
Considering the qubit encoding, this transformation is---up to an atomic qubit-flip---precisely \eqref{TGHZ}. Since a qubit-flip is nothing but an innocuous local unitary operation, the resulting state is just the desired GHZ state but in a different local (qubit-flipped) basis.

In the case of \eqref{TLC}, we set the laser polarization forming an angle $\alpha$ with the cavity axis.
That is, both components, $\pi$ polarization, with weight $\cos(\alpha)$, and $\sigma^+/\sigma^-$ polarization, with weight $\sin(\alpha)$, are now present in the polarization vector of the pump.
Therefore, the following excitations can be driven [see Fig.
\ref{Ca} $(b)$]: $\left\vert S_{1/2},m_J=\pm1/2\right\rangle \rightarrow\cos(\alpha)\left\vert P_{1/2},m_J=\pm1/2\right\rangle\mp\sin(\alpha)\left\vert P_{1/2},m_J=\mp1/2\right\rangle$.
These excitations decay via photon emission into the cavity exactly as before, yielding $|S_{1/2},m_J=\pm1/2\rangle\rightarrow\sin(\alpha)|S_{1/2},m_J=\pm1/2\rangle |\sigma_1^{\mp}\rangle\pm\cos(\alpha)|S_{1/2},m_J=\mp1/2\rangle |\sigma_1^{\pm}\rangle )$ as the total transformation for the ground states.
Once again taking into account the qubit encoding, we see that if $\alpha=\pi/4$ the latter is -- up to local unitary qubit-flips -- identical with transformation \eqref{TLC}.

\section{Implementation with $^{87}\mathrm{Rb}$}
\label{SecRb}
\begin{figure}
\begin{center}
\includegraphics[width=1\linewidth]{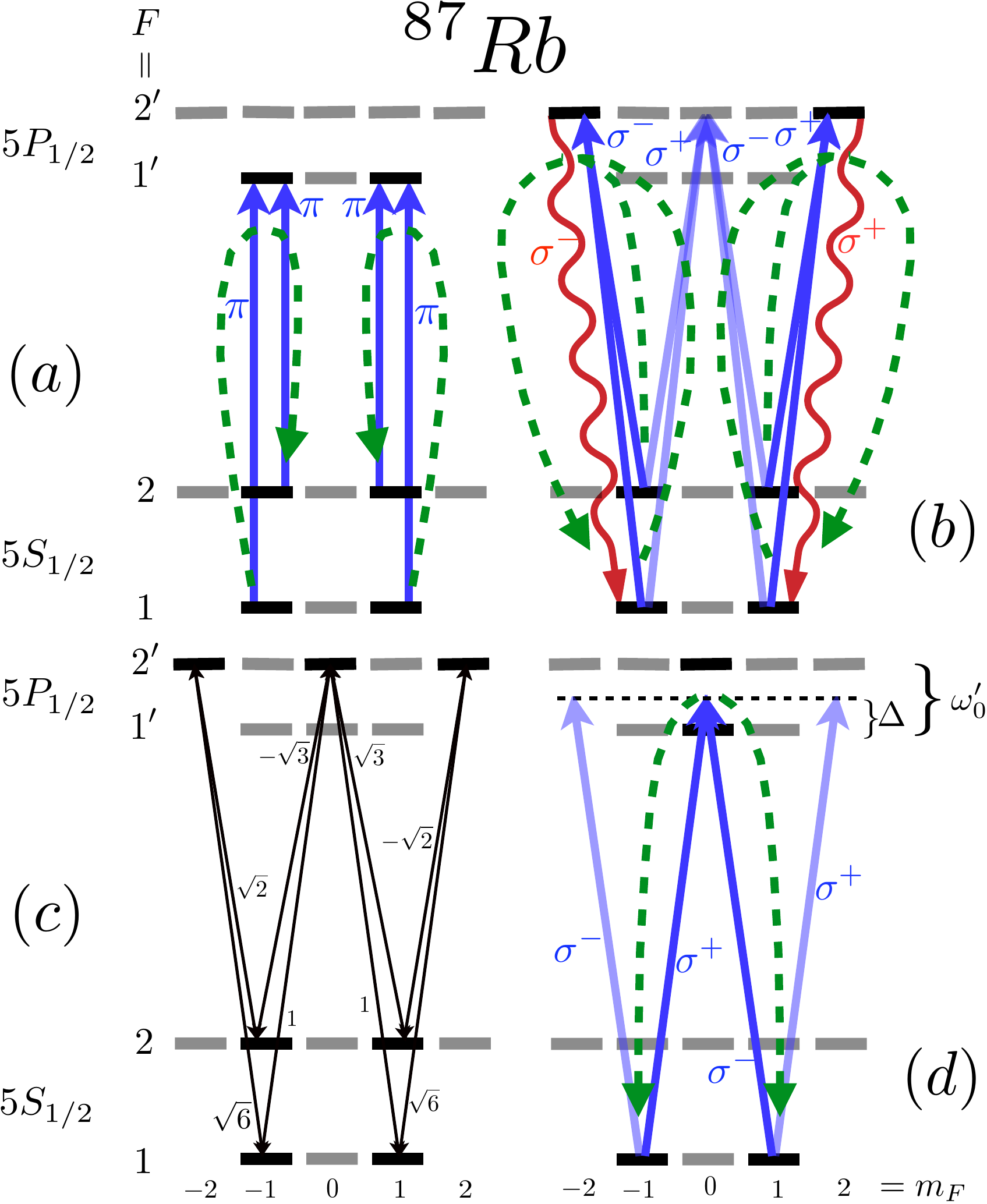}
\caption{ \label{Rb} (Color online.) Relevant hyperfine level structure of $^{87}\mathrm{Rb}$.
The cavity is resonant with the $1\leftrightarrow2'$ transition.
$(a)$ Two $\pi$-polarized lasers (blue lines), respectively in resonance with $2\leftrightarrow1'$ and $1\leftrightarrow1'$, drive a STIRAP (green dashed lines) that partially transfers population from $\ket{1,\pm1}$ to $\ket{2,\pm1}$.
$(b)$ A bichromatic $\sigma^+/\sigma^-$-polarized laser, with color components respectively in resonance with $1\leftrightarrow2'$ and $2\leftrightarrow2'$, state-selectively excites the atom to the $2'$ manifold.
State-selection is achieved exploiting the dipole matrix elements, indicated in panel $(c)$ in arbitrary units, and is such that excitations to the $\ket{2,0}$ state are blocked. Thus, photon emissions (red wavy lines) only from levels $\ket{2',\pm2}$ occur, giving rise to the effective process shown in green-dashed lines.
The composed action of $(a)$ and $(b)$ implements transformation \eqref{TGHZ}.
$(d)$ If these processes are in turn preceded by a $\pi/2$ rotation between states $\ket{1,\pm1}$, transformation \eqref{TLC} is obtained.
Such rotation is driven by a Raman transition induced by a $\sigma^+/\sigma^-$-polarized monochromatic laser $\Delta$-detuned from $1\leftrightarrow1'$.
The optimal detuning is $\Delta=\omega_0'/2$, with $\omega_0'$ being the frequency difference between $1'$ and $2'$.}
\end{center}
\end{figure}

This species possesses a rich hyperfine structure, schematically represented in Fig. \ref{Rb}.
We use sublevels $\left\vert 5^2S_{1/2},F=1,m_F=\pm 1\right\rangle\equiv\left\vert\pm\right\rangle$ of the ground-state hyperfine manifold as the atomic qubit.
The short-hand notation ``$5^2S_{1/2},F=i\rightarrow i$" and ``$5^2P_{1/2},F=i\rightarrow i^{\prime }$" is used throughout.
The cavity mode is now in resonance with the $1\leftrightarrow2^{\prime }$ transition.

We begin by the implementation of \eqref{TGHZ}, which requires two pulses. In the first one, sketched in Fig \ref{Rb} $(a)$, a two-photon Raman process resonant with the transition $1\leftrightarrow2$ partially transfers population from $|1,\pm1\rangle$ to $|2,\pm1\rangle$.
This is performed with a conventional stimulated Raman adiabatic passage (STIRAP)
, very similar to the one used in Ref. \cite{Volz}.
Two $\pi$-polarized smooth laser pulses are used.
One of them is resonant with the $2\leftrightarrow1^{\prime }$ transition and is switched on first.
The other one is resonant with $1\leftrightarrow1^{\prime }$ and is switched on (reaches peak intensity) exactly when the first one reaches peak intensity (is completely switched off).
This procedure allows for the use of a zero Raman-detuning at the same time keeping spontaneous emission negligible, for the entire evolution remains in a (adiabatically-varying) dark-state.
The STIRAP-pulse area is such that $|1,\pm1\rangle \rightarrow1/2(\mp\sqrt{3}|1,\pm1\rangle +|2,\pm1\rangle )\equiv\pm|\eta^{\pm}\rangle $.
States $|1,1\rangle $ and $|1,-1\rangle $ rotate in opposite angles because of the relative signs between the Clebsch-Gordan coefficients (not shown) of the transitions involved.

In the second step [Fig. \ref{Rb} $(b)$] a  bichromatic laser pulse excites the atom to the $2^{\prime}$ sublevels. Both laser-frequency components are $\sigma^+/\sigma^-$-polarized and have the same amplitude  (with zero optical phase, for simplicity). One of them is resonant with $1\leftrightarrow2^{\prime }$ and the other one with $2\leftrightarrow2^{\prime }$.
Taking into account the couplings shown in Fig. \ref{Rb} $(c)$ \cite{Steck}, one sees that the interaction Hamiltonian is proportional to $H=|2^{\prime },2\rangle \langle \eta^+| +|2^{\prime},-2\rangle \langle \eta^-| +|2^{\prime },0\rangle \langle \eta^{\perp}| +\text{h. c.}$, where $|\eta^{\perp}\rangle $ is a state orthogonal to both $|\eta^{+}\rangle $ and $|\eta^{-}\rangle $.
This implies that the latter are both dark states with respect to transitions to $|2^{\prime },0\rangle$ and  can therefore only be excited to $|2^{\prime },\pm2\rangle$, with the subsequent emission of a $\sigma^{\pm}$ photon. The pulse area is $\pi$, so that the excitations $|\eta^{\pm}\rangle \rightarrow|2^{\prime },\pm2\rangle$ are carried out.
Altogether, the mapping $|1,\pm1\rangle\rightarrow\pm|1,\pm1\rangle |\sigma^\pm\rangle $ is completed: transformation \eqref{TGHZ} in the chosen qubit encoding.

To end up with, transformation \eqref{TLC} requires the same two pulses just described but preceded by an extra one, sketched in Fig. \ref{Rb}$(d)$.
This consists of a monochromatic $\sigma^+/\sigma^-$-polarized Raman laser, whose frequency is exactly halfway between the $1\leftrightarrow1^{\prime }$ and $1^{\prime}\leftrightarrow2^{\prime }$ transition frequencies.
The Raman pulse area is $\pi/2$, yielding the rotation $|1,\pm1\rangle\rightarrow\frac{1}{\sqrt{2}}(\pm|1,1\rangle +|1,-1\rangle )$.
With this, the total transformation is $|1,\pm 1\rangle \rightarrow\frac{1}{\sqrt{2}}(\pm|1,1\rangle |\sigma^+\rangle -|1,-1\rangle |\sigma^-\rangle )$, the desired operation \eqref{TLC}.

In the latter, the choice of Raman detuning is not at all casual.
For arbitrary detunings $\Delta$, the effective Rabi frequency is proportional to $\frac{1}{\Delta }-\frac{1}{\Delta +\omega _{0}^{\prime}}$,
 that is, with the contributions from the virtual mediator levels $|1^{\prime },0\rangle $ and $|2^{\prime },0\rangle $ canceling out in the large-detuning limit.
On the other hand, small detunings tend to increase the risk of undesired photon-scattering events.
Nevertheless, $\Delta=\omega _{0}^{\prime}/2$ maximizes the Rabi frequency and at the same time keeps spontaneous emission within negligible levels \cite{Wright}.

\section{Technical details} 
\label{TecDet}
 To minimize the chance that more than one photon is produced per sequence, excitations can be done with fast-excitation pulses, shorter than all other relevant time scales \cite{Blinov,Darquie2005,Bochmann}. These pulses last so short  that the atom hardly has time either to decay or to transfer its excitation to the cavity before the pulse is already finished. This way the probability of multiple excitations during the same pulse can be strongly suppressed to levels as low as 0.01\%  (pulse durations of 3 ns)  \cite{Bochmann}, so as not to constitute a significant error source. A potential drawback could in principle be the broadening of the laser linewidth. In fact, the linewidth can become comparable to the hyperfine splitting between the $2^{\prime }$ and $1^{\prime}$ manifolds of $^{87}\mathrm{Rb}$, $\omega_0'\approx 814\ \mathrm{MHz}$, making it inviable to address one without addressing the other. Therefore, unwanted transitions to $\left\vert 1^{\prime },0\right\rangle $ could in principle occur in the excitation pulse of Fig. \ref{Rb} $(b)$, imposing a fundamental limitation. However, the relevant dipole matrix elements  \cite{Steck} are such that  $\left\vert \eta ^{\pm }\right\rangle$ are dark states with respect to sublevel $\left\vert 1^{\prime },0\right\rangle $ too. The protocol's performance is thus not affected by the broadening of the frequency spectrum.  

\par We have presented the scheme for $^{40}\mathrm{Ca}$ with the strong dipole transition $4S_{1/2}\leftrightarrow 4P_{1/2}$. Notice however that the same procedure can actually also be applied to weak quadrupole transitions, such as  $4S_{1/2}\leftrightarrow 3D_{3/2}$, connected by Raman processes via $4P_{1/2}$ driven jointly by the cavity and a Raman laser \cite{Keller, Barros}. In such a case, the Raman detuning can be increased so as to drastically reduce the effective spontaneous-emission rate; so that -- even though excitation pulses take longer -- extremely high overall efficiencies are attained  \cite{Barros}. 

\par In turn, for both exemplary atomic species, the extremely long coherence times (seconds) of the long-lived sublevels considered allow in principle for the production of states with several photons. For $^{87}\mathrm{Rb}$, the pulses required apart from the fast-excitation pulses -- Raman rotation and STIRAP -- can be done altogether in a few microseconds \cite{Wright, Volz}. For $^{40}\mathrm{Ca}$, even in the slowest case of Raman processes mentioned above, excitation pulses are carried out with durations of the order of 120 $\mu$s. 

\par Note also that the ``disconnection" measurement on the atom via the last emitted photon (see Appendix \ref{Transformations}) needs often not be done before the previously emitted $N$ photons arrive at destination. In such cases, the $(N+1)$th photon is measured only upon arrival and its measurement outcome is used as a feedback to post-process the previous $N$ measurement outcomes (provided of course that the atomic coherence is still intact).
For situations where the $N$-qubit photonic state must be prepared before propagation, the disconnection measurement can be done with a circularly-polarized beam-splitter mounted on a movable structure. This must be introduced in the photons' path after the $N$th photon's passage and before the last one's.
For repetition rates of up to MHz and beam waists of micrometers, a piezoelectric device coordinated with the last laser pulse can do the job.

\section{Feasibility and discussions}
\label{Feas}
 Even though   the protocol is in principle deterministic,  the overall efficiency is in practice never unity. The total probability of emission of an entangled photon pair through the cavity output observed in  Refs. \cite{Wilk,WilkPhD} is of 1.3$\%$. Nevertheless, overall efficiencies  of intracavity photon genearation and cavity photon emission of 88$\%$  and 16.7$\%$, respectively,  per photon have been demonstrated in more recent experiments \cite{Barros}. Furthermore, exhaustive simulations show that cavity photon-emission  probabilities of up to 74$\%$ per photon can be reached \cite{WilkPhD}.  Sources of inefficiencies are discussed in more detail in Appendix \ref{Photlosses}. 

\par All in all, even modest success probabilities of about 1.3$\%$ per photon pair such as the one demonstrated in Ref. \cite{Wilk} readily lie about 4 orders of magnitude above the typical efficiency ($10^{-6}$) of parametric downconversions through nonlinear crystals, used to produce entangled photon pairs  in linear-optical experiments \cite{BowmeesterPanWitnessesZhao,kieselwalther1walther2, Lu, ChenprevedelVallonececcarelli}. There, such low conversion efficiencies are overcome with pulsed sources of extremely high repetition rates and laser power. In terms of net output of entangled pairs these sources comfortably beat any cavity-based method. This changes though in the multi-partite scenario.  The creation of genuine multiphoton entangled states in linear-optical settings typically requires synchronized encounters of multiple entangled photons at beam splitters, where fusions into larger multi-photonic pieces take place. 
For any fixed pulse rate and laser power, the probability of having simultaneous pairs per shot decreases exponentially with their number. To this, one must add that every beam splitter succeeds to fuse the incoming photons only half the time, yielding an extra factor of 2 in the exponent of the net decrease. This cannot be circumvented with a tour de force increase  with $N$ of the shot repetition rate and the power, for the former increases the frequency bandwidth and the latter represents an extremely unpractical experimental overhead. The cavity-based method proposed here does not bear these particular scaling limitations and may provide a relevant alternative as one increases $N$.
 
 \par We notice further that our methods complement studies based on quantum dots \cite{Lindner}, which feature very promising scalability properties. However,  in the short term the present methods seem considerably more feasible, because -- as said -- the experimental platform they require has already repeatedly proved successful for the basic entanglement demonstrations. Finally, violations of multiqubit Bell inequalities of up to 10 photons are also viable with the  current technology. See Appendix \ref{Photlosses}. 

\section{Conclusion}
\label{Conclusion}
We have described a procedure for the creation in photonic systems of two genuine mutipartite-entangled states of fundamental importance -- the GHZ and linear cluster states. Photons are emitted by a single atom -- or ion -- inside an optical cavity. State manipulation is achieved by laser pulses with different polarizations. The relative amplitudes among the natural atomic dipole-transition elements are exploited to enhance desired couplings and block others.

\par The scheme is in principle deterministic, but in practice the overall efficiency is limited (mainly) by nonperfect intracavity photon generation and photon losses. However, cavity photon-emission efficiencies of 16.7$\%$ per photon are readily available, and values as high as close to 70$\%$ per photon seem feasible with current technology. In turn, for two-qubits, state fidelities of up to 93$\%$ are readily available too. The method appears thus as an interesting alternative to conventional linear-optical approaches for the production  of genuine multiqubit entangled states of several photons.

\par Finally, the extension of these ideas to setups with several atoms, either coupled to the same cavity or distributed in coupled-microcavity arrays, will lead to protocols for the creation of two-dimensional-cluster and general graph states.

\begin{acknowledgements}
L.A.  thanks Andreas Winter for the hospitality in Singapore, Stephan Ritter, Tatjana Wilk, Holger Specht, Gerhard Rempe, and Helena G. Barros for helpful discussions, and the Spanish ``Juan de la Cierva" program for financial support. K.L.C. and L.Y. acknowledge support from the National Research Foundation and the Ministry of Education, Singapore.
\end{acknowledgements}

\appendix
\section{GHZ and linear cluster states}
\label{Transformations}
We show here how the repeated application of transformations \eqref{GHZLC} leads to the generation of the desired multi-qubit states. Let us begin by the linear cluster state. The atom is initialized in---say---state $|\varphi _{0}\rangle \equiv |+\rangle $. Application of transformation \eqref{TLC} $N+1$ times delivers the state $\ket{{\varphi}_{N+1}}\equiv{{T_{LC}}}^{N+1} |\varphi_0\rangle\equiv\frac{1}{\sqrt{2^{N+1}}}$ $\sum_{i_1 ... i_{N+1}=\pm}(-1)^{i'_1i'_2+ ... i'_{N}i'_{N+1}+i'_{N+1}}\ket{i_{N+1}}\ket{\sigma_{N+1}^{i_{N+1}}\ ...\ \sigma_1^{i_1}}$, where we have explicitly subindexed each photon's polarization according to its order of emission. The summation goes over all possible polarization configurations.
The primed indexes in the exponent in turn denote the mapping $+^{\prime }\equiv0$ and $-^{\prime }\equiv 1$.
Also, notice that the atomic state $i_{N+1}$ is locked to the polarization $\sigma _{N+1}^{i_{N+1}}$ of the last emitted photon.
In fact, if we group the atom and the $(N+1)$-th photon together into a single effective qubit, state $|{\varphi}_{N+1}\rangle $ is already an $(N+1)$-qubit linear cluster state \cite{RausBrie}, but shared among the atom and the ${N+1}$ photons.
To disconnect the atom from the state, we simply measure it in its computational $z$ basis.
Naturally, this is most efficiently done by taking advantage of its coupling with the cavity photons:
A projective measurement on the $(N+1)$-th photon in the computational basis $\{|\sigma_{N+1}^{+}\rangle ,|\sigma _{N+1}^{-}\rangle \}$, with outcomes $\mu =0$, for $\sigma _{N+1}^{+}$, or $\mu =1$, for $\sigma _{N+1}^{-}$, disconnects the effective atom-last-photon qubit from the rest of the cluster.
The final state of the remaining $N$ photons is -- up to an innocuous, $\mu$-dependent local unitary -- nothing but the desired, fully-photonic 1D cluster state \cite{RausBrie} (omitting normalization):
\begin{eqnarray}
\label{LC}
\nonumber
\ket{{\Phi}_{N}}=\sum_{i_1 ... i_{N}=\pm}(-1)^{i'_1i'_2+ ... i'_{N-1}i'_{N}+i'_{N}\mu+\mu}\ket{\sigma_{N}^{i_{N}}...\ \sigma_1^{i_1}}.
\end{eqnarray}

\par For the case of the GHZ state, let us take the initial atom's state as $\ket{\psi_0}\equiv\frac{1}{\sqrt{2}}(\ket{+}+\ket{-})$, for instance. We apply now transformation \eqref{TGHZ} $N$ times to obtain $|{\psi_a}_N\rangle\equiv{{T_{GHZ}}_a}^N|\psi_0\rangle\equiv\frac{1}{\sqrt{2}} \big(|+\rangle\ket{\sigma_N^{+}\ ...\ \sigma_1^+}+ (-1)^{N}\ket{-}\ket{\sigma_N^{-}\ ...\ \sigma_1^-}\big)$. As above, state $|{\psi_a}_N\rangle$ is already an $N+1$-qubit atomic-photonic GHZ state.
To decouple the atom we now measure it in its $x$ basis, again via a photonic measurement. For this, we first apply transformation \eqref{TLC} once, which adds a further emitted photon with polarization locked to the atomic state as above. Next we measure this photon in its computational basis also as above. This projects the other $N$ photons onto
\begin{equation}
\label{GHZ}
|{\Psi_a}_N\rangle=(-1)^{\mu}\ket{\sigma_N^{+}\ ...\ \sigma_1^+}- (-1)^{N}\ket{\sigma_N^{-}\ ...\ \sigma_1^-},
\end{equation}
which is -- up to a $\mu$-dependent local unitary -- the desired photonic GHZ state \cite{GHZ} (normalization omitted again). Notice finally that the initial atomic state here can also be taken as $|\varphi _{0}\rangle \equiv |+\rangle$. In this case the protocol is the same except for the first of the $N+1$ required transformations, which is of the type \eqref{TLC} instead of \eqref{TGHZ}. The resulting state is -- up to a minus sign -- also given by \eqref{GHZ}.

\section{Efficiencies and fidelities}
\label{Photlosses}

\par The main sources of inefficiencies in the emission of photons from the cavity are nonperfect intracavity photon generation and photon losses. The former is essentially due to atomic motion (which introduces uncontrolled variations in the atom-cavity coupling) and imperfections in the pump. The latter is mostly dominated by atomic spontaneous emission and absorption or scattering from the cavity mirrors.  One alternative to dominate atomic motion is to consider ionized specimens and exploit the strong electrical confinement available in ion traps.
For $^{40}\mathrm{Ca}^+$, overall efficiencies per photon of intracavity photon generation and cavity photon emission of 88$\%$  and 16.7$\%$, respectively,  have been recently demonstrated  \cite{Barros}. There, advantage was also taken of the reduced effective spontaneous emission rate due to large Raman detunings and the use of weak transitions. Cavity-photon loss was mostly due to mirror scattering. Another possibility for tight confinement is strong cooling and optical dipole traps \cite{Weber}. Indeed, for $^{87}\mathrm{Rb}$, with appropriate cavity-atom and cavity-pump detunings, simulations  show that, for realistic cavity-atom couplings such as $g/2\pi=6.7$ MHz, the overall probability of photonic emission from the cavity can be enhanced up to 74$\%$ per photon when the atomic motion is neglected (see Chapt. 3 of Ref. \cite{WilkPhD}). These simulations take into account spontaneous emission, undesired off-resonant excitations to other levels and magnetic fields, and yield total photon losses due to atomic spontaneous emission below 15$\%$. It is interesting to notice that such high efficiency is above the threshold---50$\%$---of loss-tolerant photonic one-way quantum computing \cite{Michael}. 

\par It is also important however to keep in mind that detection efficiencies (including non-perfect mode-matching into the fiber, transmission losses through the fiber and detector efficiencies) are usually no more than 30$\%$. Nevertheless, non-perfect detection is inherent to any photonic-state manipulation scheme and is therefore not a figure of merit for the efficiency of photonic-state generation schemes, such as the one proposed here.

\par Finally, we notice that fidelities of 86$\%$, 87$\%$ and 93$\%$ for two-qubit maximally-entangled states have been reported (see \cite{Wilk, WilkPhD}, \cite{Blinov} and \cite{Weber}, respectively). In turn, the minimal fidelities required for the demonstration of genuine multipartite entanglement using graph-state entanglement witnesses, or for the violation of genuine multipartite Bell inequalities, go from 75$\%$, for 3 qubits, to approximately 53$\%$ and 35$\%$, for GHZ and linear cluster states, respectively, for 10 qubits \cite{Otfried}. Thus the methods proposed here open a realistic venue for photonic multiqubit entanglement and non-locality experiments with high efficiency.


\end{document}